\documentclass[a4paper]{jpconf}
\usepackage{graphicx}
\usepackage{amsmath}
\usepackage{accents}
\newlength{\dhatheight}
\newcommand{\doublehat}[1]{%
    \settoheight{\dhatheight}{\ensuremath{\hat{#1}}}%
    \addtolength{\dhatheight}{-0.35ex}%
    \hat{\vphantom{\rule{1pt}{\dhatheight}}%
    \smash{\hat{#1}}}}
\usepackage{tikz}
\usetikzlibrary{backgrounds}
\usetikzlibrary{arrows,shapes}
\usetikzlibrary{tikzmark}
\usetikzlibrary{calc}

\usepackage{amsmath}
\usepackage{amsthm}
\usepackage{amssymb}
\usepackage{mathtools}
\usepackage{wrapfig}
\usepackage{comment}
\usepackage{hyperref}
\usepackage{float}
\newcommand{\weights}{$\textcolor[rgb]{0,0.46,1}{\varphi }$}

\begin{document}
\title{\texttt{neos}: End-to-End-Optimised Summary Statistics for High Energy Physics}

\author{Nathan Simpson$^{1}$ and Lukas Heinrich$^{2}$}

\address{$^1$Lund University}
\address{$^2$Technical University of Munich}

\ead{n.s@cern.ch, lukas.heinrich@cern.ch}

\newcommand{\CLs}{$\mathrm{CL_s}$}
\begin{abstract}
The advent of deep learning has yielded powerful tools to automatically compute gradients of computations. This is because training a neural network equates to iteratively updating its parameters using gradient descent to find the minimum of a loss function. Deep learning is then a subset of a broader paradigm; a workflow with free parameters that is end-to-end optimisable, provided one can keep track of the gradients all the way through.

This work introduces \texttt{neos}: an example implementation following this paradigm of a \textit{fully differentiable high-energy physics workflow}, capable of optimising a learnable summary statistic with respect to the expected sensitivity of an analysis. Doing this results in an optimisation process that is aware of the modelling and treatment of systematic uncertainties.
\end{abstract}

\section{Introduction}
\textbf{Summary statistics} are typically low-dimensional quantities
calculated from high-dimensional data, often for the purpose of making
any following inference stage easier to compute. They find
widespread use in high-energy physics (HEP) data analysis for this reason, since HEP data, e.g. that produced at the Large Hadron Collider, is characterised by both its high dimensionality ($\mathcal{O}(10^8)$ detector read-out channels) as well as the intractability of computing its probability $p(x|\theta)$ under a given physics hypothesis $\theta$. A notable example of such a quantity would be the invariant mass of the Higgs boson, which was used in its discovery in 2012 \cite{2012}.

Thanks to the presence of high-quality physics simulators, effective machine-learning based summary statistics can be formed using simulated data, typically being trained using classification-driven objectives to maximally distinguish new physics from background processes. While these objectives can be shown to produce optimal results in simplified hypothesis testing settings, the presence of systematic uncertainties in the data may cause classification objectives to become misaligned with physics objectives, e.g. the expected sensitivity to a signal. This means that algorithms trained in this way may under-perform when integrated into the actual physics analysis pipeline, motivating the search for objectives that can better adapt to data with high systematic uncertainty.

To combat this, we propose to optimise the summary statistic directly on physics objectives by formulating an \textit{end-to-end-differentiable analysis pipeline}, including density estimation of the summary statistic with histograms, and statistical inference using standard HEP prescriptions. The training can then based on physics-oriented metrics, such as the expected sensitivity, which is based on the profile likelihood ratio: a quantity capable of capturing systematic effects.

\subsection{Related work}
The most similar work to \texttt{neos} is INFERNO \cite{inferno}, which also targets the optimisation of a summary statistic with respect to an inference-aware loss function. We compare our approach to INFERNO below both qualitatively and quantitatively.
Other attempts to incorporate robustness to systematic uncertainties include: directly parameterising the neural network in the relevant nuisance parameters that model systematic uncertainties \cite{uncert}, and including an adversarial term in the loss that penalises dependence on these nuisance parameters \cite{pivot}. 

\section{Making HEP Analysis Differentiable}

Given a pre-filtered dataset, an analysis pipeline in HEP involves the following stages:
\begin{enumerate}
    \item Construction of a learnable 1-D summary statistic from data (with parameters $\varphi$)
    \item Binning of the summary statistic, e.g. through a histogram
    \item Statistical model building, using the summary statistic as a template
    \item Calculation of a test statistic, used to perform a frequentist hypothesis test of signal versus background
    \item A $p$-value (or \CLs{}\footnote{\CLs{} is a modification of the p-value that protects
against rejecting the null hypothesis when the test is not sensitive to
the alternative hypothesis (e.g.~through largely overlapping test
statistic distributions).} value) resulting from that hypothesis test, used to characterise the sensitivity of the analysis
\end{enumerate}

We can express this workflow as a direct function of the input dataset $\mathcal{D}$ and observable parameters $\varphi$:

\tikzset{every picture/.style={line width=0.75pt}}
\begin{figure}
\centering

\begin{tikzpicture}[x=0.75pt,y=0.75pt,yscale=-1,xscale=1]

\draw  [draw opacity=0][fill={rgb, 255:red, 246; green, 199; blue, 120 }  ,fill opacity=1 ] (562,106) -- (576.67,114) -- (562,114) -- cycle ;
\draw [color={rgb, 255:red, 128; green, 128; blue, 128 }  ,draw opacity=1 ] (535,114.13) -- (596,114.13)(541.1,83) -- (541.1,117.59) (589,109.13) -- (596,114.13) -- (589,119.13) (536.1,90) -- (541.1,83) -- (546.1,90)  ;
\draw [color={rgb, 255:red, 69; green, 213; blue, 218 }  ,draw opacity=1 ][line width=1.5]    (542,92) .. controls (587,93) and (577,115) .. (592,112) ;
\draw [color={rgb, 255:red, 245; green, 166; blue, 35 }  ,draw opacity=1 ][line width=1.5]    (542,92) .. controls (570,106) and (568,116) .. (583,113) ;

\draw  [draw opacity=0][fill={rgb, 255:red, 21; green, 222; blue, 208 }  ,fill opacity=1 ] (256.84,64.01) -- (267.09,64.01) -- (267.09,79.96) -- (256.84,79.96) -- cycle ;
\draw  [draw opacity=0][fill={rgb, 255:red, 21; green, 222; blue, 208 }  ,fill opacity=1 ] (267.09,72.38) -- (277.6,72.38) -- (277.6,79.96) -- (267.09,79.96) -- cycle ;
\draw  [draw opacity=0][fill={rgb, 255:red, 21; green, 222; blue, 208 }  ,fill opacity=1 ] (277.6,67.2) -- (288.62,67.2) -- (288.62,79.96) -- (277.6,79.96) -- cycle ;

\draw  [draw opacity=0][fill={rgb, 255:red, 245; green, 166; blue, 35 }  ,fill opacity=1 ] (255.08,103) -- (265.67,103) -- (265.67,109.5) -- (255.08,109.5) -- cycle ;
\draw  [draw opacity=0][fill={rgb, 255:red, 245; green, 166; blue, 35 }  ,fill opacity=1 ] (265.61,98.86) -- (277.15,98.86) -- (277.15,109.5) -- (265.61,109.5) -- cycle ;
\draw  [draw opacity=0][fill={rgb, 255:red, 245; green, 166; blue, 35 }  ,fill opacity=1 ] (276.41,86.46) -- (287.74,86.46) -- (287.74,109.5) -- (276.41,109.5) -- cycle ;
\draw  [draw opacity=0][fill={rgb, 255:red, 208; green, 2; blue, 27 }  ,fill opacity=1 ] (297.45,71.69) -- (308.93,71.69) -- (308.93,79.96) -- (297.45,79.96) -- cycle ;
\draw  [draw opacity=0][fill={rgb, 255:red, 208; green, 2; blue, 27 }  ,fill opacity=1 ] (307.99,69.91) -- (319.52,69.91) -- (319.52,79.96) -- (307.99,79.96) -- cycle ;
\draw  [draw opacity=0][fill={rgb, 255:red, 208; green, 2; blue, 27 }  ,fill opacity=1 ] (318.78,55.14) -- (331,55.14) -- (331,79.96) -- (318.78,79.96) -- cycle ;

\draw  [draw opacity=0][fill={rgb, 255:red, 248; green, 231; blue, 28 }  ,fill opacity=1 ] (297.45,103.59) -- (308.93,103.59) -- (308.93,109.5) -- (297.45,109.5) -- cycle ;
\draw  [draw opacity=0][fill={rgb, 255:red, 248; green, 231; blue, 28 }  ,fill opacity=1 ] (307.99,102.33) -- (319.52,102.33) -- (319.52,109.5) -- (307.99,109.5) -- cycle ;
\draw  [draw opacity=0][fill={rgb, 255:red, 248; green, 231; blue, 28 }  ,fill opacity=1 ] (318.78,91.77) -- (331,91.77) -- (331,109.5) -- (318.78,109.5) -- cycle ;

\draw  [fill={rgb, 255:red, 80; green, 227; blue, 212 }  ,fill opacity=1 ] (26.4,56.5) -- (50.75,56.5) -- (50.75,73.06) .. controls (35.53,73.06) and (38.57,79.04) .. (26.4,75.17) -- cycle ; \draw  [fill={rgb, 255:red, 80; green, 227; blue, 212 }  ,fill opacity=1 ] (23.36,59.01) -- (47.7,59.01) -- (47.7,75.57) .. controls (32.49,75.57) and (35.53,81.55) .. (23.36,77.68) -- cycle ; \draw  [fill={rgb, 255:red, 80; green, 227; blue, 212 }  ,fill opacity=1 ] (20.32,61.52) -- (44.66,61.52) -- (44.66,78.08) .. controls (29.44,78.08) and (32.49,84.06) .. (20.32,80.19) -- cycle ;
\draw  [fill={rgb, 255:red, 245; green, 166; blue, 35 }  ,fill opacity=1 ] (26.4,87.5) -- (50.75,87.5) -- (50.75,104.07) .. controls (35.53,104.07) and (38.57,110.04) .. (26.4,106.18) -- cycle ; \draw  [fill={rgb, 255:red, 245; green, 166; blue, 35 }  ,fill opacity=1 ] (23.36,90.01) -- (47.7,90.01) -- (47.7,106.58) .. controls (32.49,106.58) and (35.53,112.55) .. (23.36,108.69) -- cycle ; \draw  [fill={rgb, 255:red, 245; green, 166; blue, 35 }  ,fill opacity=1 ] (20.32,92.52) -- (44.66,92.52) -- (44.66,109.09) .. controls (29.44,109.09) and (32.49,115.06) .. (20.32,111.19) -- cycle ;
\draw  [fill={rgb, 255:red, 208; green, 2; blue, 27 }  ,fill opacity=1 ] (64.66,56.5) -- (89,56.5) -- (89,73.06) .. controls (73.79,73.06) and (76.83,79.04) .. (64.66,75.17) -- cycle ; \draw  [fill={rgb, 255:red, 208; green, 2; blue, 27 }  ,fill opacity=1 ] (61.61,59.01) -- (85.96,59.01) -- (85.96,75.57) .. controls (70.74,75.57) and (73.79,81.55) .. (61.61,77.68) -- cycle ; \draw  [fill={rgb, 255:red, 208; green, 2; blue, 27 }  ,fill opacity=1 ] (58.57,61.52) -- (82.91,61.52) -- (82.91,78.08) .. controls (67.7,78.08) and (70.74,84.06) .. (58.57,80.19) -- cycle ;
\draw  [fill={rgb, 255:red, 248; green, 231; blue, 28 }  ,fill opacity=1 ] (64.66,87.5) -- (89,87.5) -- (89,104.07) .. controls (73.79,104.07) and (76.83,110.04) .. (64.66,106.18) -- cycle ; \draw  [fill={rgb, 255:red, 248; green, 231; blue, 28 }  ,fill opacity=1 ] (61.61,90.01) -- (85.96,90.01) -- (85.96,106.58) .. controls (70.74,106.58) and (73.79,112.55) .. (61.61,108.69) -- cycle ; \draw  [fill={rgb, 255:red, 248; green, 231; blue, 28 }  ,fill opacity=1 ] (58.57,92.52) -- (82.91,92.52) -- (82.91,109.09) .. controls (67.7,109.09) and (70.74,115.06) .. (58.57,111.19) -- cycle ;

\draw  [color={rgb, 255:red, 53; green, 184; blue, 254 }  ,draw opacity=1 ][fill={rgb, 255:red, 163; green, 205; blue, 254 }  ,fill opacity=1 ] (142,64) .. controls (142,60.13) and (145.13,57) .. (149,57) .. controls (152.87,57) and (156,60.13) .. (156,64) .. controls (156,67.87) and (152.87,71) .. (149,71) .. controls (145.13,71) and (142,67.87) .. (142,64) -- cycle ;
\draw  [color={rgb, 255:red, 53; green, 184; blue, 254 }  ,draw opacity=1 ][fill={rgb, 255:red, 163; green, 205; blue, 254 }  ,fill opacity=1 ] (142,84) .. controls (142,80.13) and (145.13,77) .. (149,77) .. controls (152.87,77) and (156,80.13) .. (156,84) .. controls (156,87.87) and (152.87,91) .. (149,91) .. controls (145.13,91) and (142,87.87) .. (142,84) -- cycle ;
\draw  [color={rgb, 255:red, 53; green, 184; blue, 254 }  ,draw opacity=1 ][fill={rgb, 255:red, 163; green, 205; blue, 254 }  ,fill opacity=1 ] (142,104) .. controls (142,100.13) and (145.13,97) .. (149,97) .. controls (152.87,97) and (156,100.13) .. (156,104) .. controls (156,107.87) and (152.87,111) .. (149,111) .. controls (145.13,111) and (142,107.87) .. (142,104) -- cycle ;
\draw  [color={rgb, 255:red, 53; green, 184; blue, 254 }  ,draw opacity=1 ][fill={rgb, 255:red, 163; green, 205; blue, 254 }  ,fill opacity=1 ] (142,125) .. controls (142,121.13) and (145.13,118) .. (149,118) .. controls (152.87,118) and (156,121.13) .. (156,125) .. controls (156,128.87) and (152.87,132) .. (149,132) .. controls (145.13,132) and (142,128.87) .. (142,125) -- cycle ;
\draw  [color={rgb, 255:red, 53; green, 184; blue, 254 }  ,draw opacity=1 ][fill={rgb, 255:red, 163; green, 205; blue, 254 }  ,fill opacity=1 ] (164,76) .. controls (164,72.13) and (167.13,69) .. (171,69) .. controls (174.87,69) and (178,72.13) .. (178,76) .. controls (178,79.87) and (174.87,83) .. (171,83) .. controls (167.13,83) and (164,79.87) .. (164,76) -- cycle ;
\draw  [color={rgb, 255:red, 53; green, 184; blue, 254 }  ,draw opacity=1 ][fill={rgb, 255:red, 163; green, 205; blue, 254 }  ,fill opacity=1 ] (164,96) .. controls (164,92.13) and (167.13,89) .. (171,89) .. controls (174.87,89) and (178,92.13) .. (178,96) .. controls (178,99.87) and (174.87,103) .. (171,103) .. controls (167.13,103) and (164,99.87) .. (164,96) -- cycle ;
\draw  [color={rgb, 255:red, 53; green, 184; blue, 254 }  ,draw opacity=1 ][fill={rgb, 255:red, 163; green, 205; blue, 254 }  ,fill opacity=1 ] (164,116) .. controls (164,112.13) and (167.13,109) .. (171,109) .. controls (174.87,109) and (178,112.13) .. (178,116) .. controls (178,119.87) and (174.87,123) .. (171,123) .. controls (167.13,123) and (164,119.87) .. (164,116) -- cycle ;
\draw  [color={rgb, 255:red, 53; green, 184; blue, 254 }  ,draw opacity=1 ][fill={rgb, 255:red, 163; green, 205; blue, 254 }  ,fill opacity=1 ] (187,96) .. controls (187,92.13) and (190.13,89) .. (194,89) .. controls (197.87,89) and (201,92.13) .. (201,96) .. controls (201,99.87) and (197.87,103) .. (194,103) .. controls (190.13,103) and (187,99.87) .. (187,96) -- cycle ;
\draw [color={rgb, 255:red, 128; green, 128; blue, 128 }  ,draw opacity=1 ]   (156,64) -- (164,76) ;
\draw [color={rgb, 255:red, 128; green, 128; blue, 128 }  ,draw opacity=1 ]   (156,64) -- (164,96) ;
\draw [color={rgb, 255:red, 128; green, 128; blue, 128 }  ,draw opacity=1 ]   (156,64) -- (164,116) ;
\draw [color={rgb, 255:red, 128; green, 128; blue, 128 }  ,draw opacity=1 ]   (164,76) -- (156,84) ;
\draw [color={rgb, 255:red, 128; green, 128; blue, 128 }  ,draw opacity=1 ]   (156,84) -- (164,96) ;
\draw [color={rgb, 255:red, 128; green, 128; blue, 128 }  ,draw opacity=1 ]   (156,84) -- (164,116) ;
\draw [color={rgb, 255:red, 128; green, 128; blue, 128 }  ,draw opacity=1 ]   (156,104) -- (164,76) ;
\draw [color={rgb, 255:red, 128; green, 128; blue, 128 }  ,draw opacity=1 ]   (156,104) -- (164,96) ;
\draw [color={rgb, 255:red, 128; green, 128; blue, 128 }  ,draw opacity=1 ]   (164,116) -- (156,104) ;
\draw [color={rgb, 255:red, 128; green, 128; blue, 128 }  ,draw opacity=1 ]   (164,76) -- (156,125) ;
\draw [color={rgb, 255:red, 128; green, 128; blue, 128 }  ,draw opacity=1 ]   (164,96) -- (157,123) ;
\draw [color={rgb, 255:red, 128; green, 128; blue, 128 }  ,draw opacity=1 ]   (164,116) -- (157,123) ;
\draw [color={rgb, 255:red, 128; green, 128; blue, 128 }  ,draw opacity=1 ]   (178,76) -- (187,96) ;
\draw [color={rgb, 255:red, 128; green, 128; blue, 128 }  ,draw opacity=1 ]   (178,96) -- (187,96) ;
\draw [color={rgb, 255:red, 128; green, 128; blue, 128 }  ,draw opacity=1 ]   (178,116) -- (187,96) ;

\draw   (2,64.81) .. controls (2,56.06) and (9.09,48.96) .. (17.84,48.96) -- (91.92,48.96) .. controls (100.67,48.96) and (107.77,56.06) .. (107.77,64.81) -- (107.77,123.66) .. controls (107.77,132.41) and (100.67,139.5) .. (91.92,139.5) -- (17.84,139.5) .. controls (9.09,139.5) and (2,132.41) .. (2,123.66) -- cycle ;
\draw  [color={rgb, 255:red, 4; green, 99; blue, 210 }  ,draw opacity=1 ] (136.5,65.11) .. controls (136.5,57.62) and (142.57,51.55) .. (150.06,51.55) -- (200.44,51.55) .. controls (207.93,51.55) and (214,57.62) .. (214,65.11) -- (214,126.89) .. controls (214,134.38) and (207.93,140.45) .. (200.44,140.45) -- (150.06,140.45) .. controls (142.57,140.45) and (136.5,134.38) .. (136.5,126.89) -- cycle ;
\draw  [color={rgb, 255:red, 144; green, 19; blue, 254 }  ,draw opacity=1 ] (240,65.75) .. controls (240,57.05) and (247.05,50) .. (255.75,50) -- (332.25,50) .. controls (340.95,50) and (348,57.05) .. (348,65.75) -- (348,124.25) .. controls (348,132.95) and (340.95,140) .. (332.25,140) -- (255.75,140) .. controls (247.05,140) and (240,132.95) .. (240,124.25) -- cycle ;
\draw  [color={rgb, 255:red, 126; green, 211; blue, 33 }  ,draw opacity=1 ] (372,84.9) .. controls (372,82.19) and (374.19,80) .. (376.9,80) -- (439.1,80) .. controls (441.81,80) and (444,82.19) .. (444,84.9) -- (444,103.1) .. controls (444,105.81) and (441.81,108) .. (439.1,108) -- (376.9,108) .. controls (374.19,108) and (372,105.81) .. (372,103.1) -- cycle ;
\draw  [color={rgb, 255:red, 68; green, 202; blue, 212 }  ,draw opacity=1 ] (464,84.9) .. controls (464,82.19) and (466.19,80) .. (468.9,80) -- (506.1,80) .. controls (508.81,80) and (511,82.19) .. (511,84.9) -- (511,103.1) .. controls (511,105.81) and (508.81,108) .. (506.1,108) -- (468.9,108) .. controls (466.19,108) and (464,105.81) .. (464,103.1) -- cycle ;
\draw    (109,97) -- (134,97) ;
\draw [shift={(136,97)}, rotate = 180] [color={rgb, 255:red, 0; green, 0; blue, 0 }  ][line width=0.75]    (10.93,-3.29) .. controls (6.95,-1.4) and (3.31,-0.3) .. (0,0) .. controls (3.31,0.3) and (6.95,1.4) .. (10.93,3.29)   ;
\draw    (214,97) -- (238,97) ;
\draw [shift={(240,97)}, rotate = 180] [color={rgb, 255:red, 0; green, 0; blue, 0 }  ][line width=0.75]    (10.93,-3.29) .. controls (6.95,-1.4) and (3.31,-0.3) .. (0,0) .. controls (3.31,0.3) and (6.95,1.4) .. (10.93,3.29)   ;
\draw    (444,96) -- (455,96) -- (461,96) ;
\draw [shift={(463,96)}, rotate = 180] [color={rgb, 255:red, 0; green, 0; blue, 0 }  ][line width=0.75]    (10.93,-3.29) .. controls (6.95,-1.4) and (3.31,-0.3) .. (0,0) .. controls (3.31,0.3) and (6.95,1.4) .. (10.93,3.29)   ;
\draw    (348,96) -- (370,96) ;
\draw [shift={(372,96)}, rotate = 180] [color={rgb, 255:red, 0; green, 0; blue, 0 }  ][line width=0.75]    (10.93,-3.29) .. controls (6.95,-1.4) and (3.31,-0.3) .. (0,0) .. controls (3.31,0.3) and (6.95,1.4) .. (10.93,3.29)   ;
\draw  [color={rgb, 255:red, 226; green, 2; blue, 174 }  ,draw opacity=1 ] (532,74.74) .. controls (532,69.08) and (536.58,64.5) .. (542.24,64.5) -- (589.76,64.5) .. controls (595.42,64.5) and (600,69.08) .. (600,74.74) -- (600,112.76) .. controls (600,118.42) and (595.42,123) .. (589.76,123) -- (542.24,123) .. controls (536.58,123) and (532,118.42) .. (532,112.76) -- cycle ;
\draw    (512,96) -- (530,96) ;
\draw [shift={(532,96)}, rotate = 180] [color={rgb, 255:red, 0; green, 0; blue, 0 }  ][line width=0.75]    (10.93,-3.29) .. controls (6.95,-1.4) and (3.31,-0.3) .. (0,0) .. controls (3.31,0.3) and (6.95,1.4) .. (10.93,3.29)   ;
\draw    (565.76,64.5) -- (565.76,40.5) ;
\draw [shift={(565.76,38.5)}, rotate = 90] [color={rgb, 255:red, 0; green, 0; blue, 0 }  ][line width=0.75]    (10.93,-3.29) .. controls (6.95,-1.4) and (3.31,-0.3) .. (0,0) .. controls (3.31,0.3) and (6.95,1.4) .. (10.93,3.29)   ;
\draw [color={rgb, 255:red, 74; green, 74; blue, 74 }  ,draw opacity=1 ] [dash pattern={on 3.75pt off 3pt on 7.5pt off 1.5pt}]  (171.68,48.71) .. controls (220.07,19.78) and (489.24,2.49) .. (550,26.5) ;
\draw [shift={(169,50.5)}, rotate = 323.13] [fill={rgb, 255:red, 74; green, 74; blue, 74 }  ,fill opacity=1 ][line width=0.08]  [draw opacity=0] (8.93,-4.29) -- (0,0) -- (8.93,4.29) -- cycle    ;
\draw [color={rgb, 255:red, 74; green, 74; blue, 74 }  ,draw opacity=1 ]   (562,87) -- (562,114) ;

\draw (467,85.3) node [anchor=north west][inner sep=0.75pt]    {$\textcolor[rgb]{0,0,0}{q_{\textcolor[rgb]{0,0,0}{\mu }}( x) \ }$};
\draw (7,119.4) node [anchor=north west][inner sep=0.75pt]  [font=\footnotesize]  {$\textcolor[rgb]{0,0,0}{\{}\textcolor[rgb]{0.31,0.89,0.87}{s} ,\textcolor[rgb]{0.96,0.65,0.14}{b} ,\textcolor[rgb]{0.73,0,0.09}{b}\mathrm{\textcolor[rgb]{0.73,0,0.09}{_{up}}} ,\textcolor[rgb]{0.93,0.86,0}{b}\textcolor[rgb]{0.93,0.86,0}{_{\mathrm{down}}}\textcolor[rgb]{0,0,0}{\}}\textcolor[rgb]{0,0,0}{_{i}} \ $};
\draw (3.37,103.63) node [anchor=north west][inner sep=0.75pt]  [font=\footnotesize]  {$ \begin{array}{l}
\end{array}$};
\draw (31,143.4) node [anchor=north west][inner sep=0.75pt]  [font=\footnotesize,color={rgb, 255:red, 0; green, 0; blue, 0 }  ,opacity=1 ]  {$\mathrm{data} \ ( d_{i})$};
\draw (134,142.4) node [anchor=north west][inner sep=0.75pt]  [font=\footnotesize,color={rgb, 255:red, 0; green, 117; blue, 255 }  ,opacity=1 ]  {$\mathrm{neural\ network}$};
\draw (234,142.4) node [anchor=north west][inner sep=0.75pt]  [font=\footnotesize,color={rgb, 255:red, 144; green, 19; blue, 254 }  ,opacity=1 ]  {$\mathrm{binned\ summary\ stat.}$};
\draw (247.75,119.4) node [anchor=north west][inner sep=0.75pt]  [font=\footnotesize,color={rgb, 255:red, 0; green, 0; blue, 0 }  ,opacity=1 ]  {$\textcolor[rgb]{0,0,0}{h_{i} =\mathrm{hist}( f_{\textcolor[rgb]{0,0.46,1}{\varphi }}( d_{i}))}$};
\draw (462.9,112.4) node [anchor=north west][inner sep=0.75pt]  [font=\footnotesize,color={rgb, 255:red, 39; green, 209; blue, 246 }  ,opacity=1 ]  {$ \begin{array}{l}
\mathrm{test\ }\\
\mathrm{statistic}
\end{array}$};
\draw (532.9,125.4) node [anchor=north west][inner sep=0.75pt]  [font=\footnotesize,color={rgb, 255:red, 224; green, 16; blue, 181 }  ,opacity=1 ]  {$ \begin{array}{l}
\mathrm{hypothesis}\\
\mathrm{test}
\end{array}$};
\draw (373.9,112.4) node [anchor=north west][inner sep=0.75pt]  [font=\footnotesize,color={rgb, 255:red, 126; green, 211; blue, 33 }  ,opacity=1 ]  {$ \begin{array}{l}
\mathrm{likelihood\ }\\
\mathrm{model}
\end{array}$};
\draw (374,84.4) node [anchor=north west][inner sep=0.75pt]    {$\textcolor[rgb]{0,0,0}{p_{\mathbf{\textcolor[rgb]{0,0,0}{h}}}( x|\mu ,\theta )}$};
\draw (554.9,21.4) node [anchor=north west][inner sep=0.75pt]  [font=\footnotesize,color={rgb, 255:red, 224; green, 16; blue, 181 }  ,opacity=1 ]  {$\mathrm{\textcolor[rgb]{0,0,0}{CL_{s}}}$};
\draw (58.9,22.4) node [anchor=north west][inner sep=0.75pt]  [font=\footnotesize,color={rgb, 255:red, 126; green, 211; blue, 33 }  ,opacity=1 ]  {$\mathrm{\textcolor[rgb]{0.29,0.29,0.29}{update\ }\textcolor[rgb]{0,0.46,1}{\varphi }\textcolor[rgb]{0.29,0.29,0.29}{\ with}}\textcolor[rgb]{0.29,0.29,0.29}{\ \frac{\partial \mathrm{\textcolor[rgb]{0.29,0.29,0.29}{CL_{s}}}}{\partial \textcolor[rgb]{0,0.46,1}{\varphi }}}$};
\draw (378.9,21.4) node [anchor=north west][inner sep=0.75pt]  [font=\footnotesize,color={rgb, 255:red, 126; green, 211; blue, 33 }  ,opacity=1 ]  {$\mathrm{\textcolor[rgb]{0.29,0.29,0.29}{backward\ pass}}$};
\draw (556.1,69.53) node [anchor=north west][inner sep=0.75pt]  [font=\scriptsize]  {$\textcolor[rgb]{0,0,0}{q_{\textcolor[rgb]{0,0,0}{\mu }}( x_{\textcolor[rgb]{0,0,0}{obs}}) \ }$};
\draw (176,120.4) node [anchor=north west][inner sep=0.75pt]  [font=\footnotesize,color={rgb, 255:red, 0; green, 117; blue, 255 }  ,opacity=1 ]  {$\textcolor[rgb]{0,0,0}{f_{\textcolor[rgb]{0,0.46,1}{\varphi }}( d_{i})}$};

\end{tikzpicture}

\caption{The pipeline for \texttt{neos}. The dashed line indicating the backward pass involves updating the weights \weights{} of the neural network via gradient descent.} \label{fig:pipe}
\end{figure}
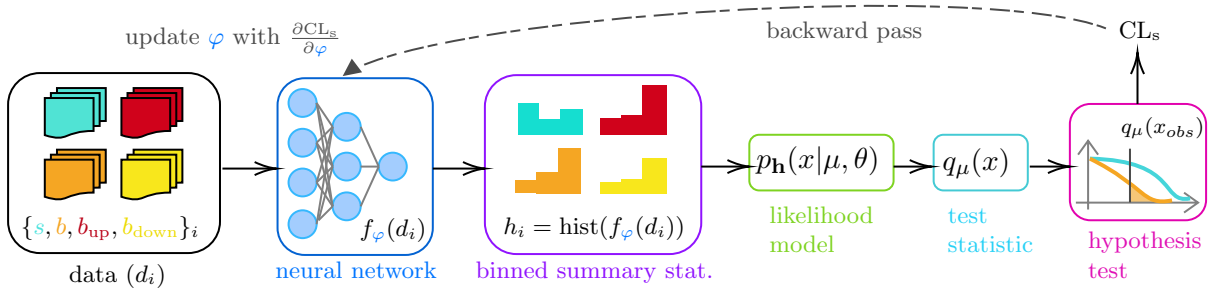

\begin{equation}\label{eq:decomp}
    \mathrm{CL_s} = f(\mathcal{D},\varphi) = (f_{\mathrm{sensitivity}} \circ f_{\mathrm{test\,stat}} \circ f_{\mathrm{likelihood}}  \circ f_{\mathrm{histogram}}  \circ f_{\mathrm{observable}})(\mathcal{D},\varphi).
\end{equation}

In the common case where $ f_{\mathrm{observable}}$ is a neural network, it seems possible to optimise this composition end-to-end, i.e. \textit{train the network to directly optimise the analysis sensitivity}. This is exactly the task that \texttt{neos} sets out to accomplish, with a full workflow detailed in \autoref{fig:pipe}. To train this network by gradient descent, our choice of $\mathrm{loss} = \mathrm{CL_s}$ requires us to be able to calculate $\partial \,\mathrm{CL_s} / \partial \varphi$. However, this is a stronger condition than it seems, and in fact necessitates the differentiablility of \textit{each individual analysis step} via the chain rule applied to \autoref{eq:decomp}.

Owing to the fact that neural networks are already differentiable, the last term $\partial f_{\mathrm{observable}}/{{\partial \varphi}}$ isn't an issue, but none of the rest of the steps are differentiable by default. The following sections detail solutions for calculating the gradient of each intermediate step.

\subsection{Binned density estimation (histograms)}

Histograms are discontinuous by nature. They are defined for 1-D data as a set of two quantities: intervals (or \textit{bins}) over the domain of that data, and counts of the number of data points that fall into each bin. For small changes in the underlying data distribution, bin counts will either remain static, or jump in integer intervals as data migrate between bins, both of which result in ill-defined gradients. We address this inherent non-differentiability through implementing a differentiable surrogate: a histogram based on a \textit{kernel density estimate} (KDE), which produces a non-parametric density $p(t)$ estimate based on samples $t_i$ and a kernel function $K(t,t_i) = K(t-t_i)$, with the full density given as $p(t) = 1/n\sum_i K(t,t_i)$. Normally, a popular kernel function choice is the standard normal distribution, which comes with a parameter called the \textbf{bandwidth} that affects the smoothness of the resulting density estimate.

Coming back to gradients: in our case, the data $t_i$ we construct the density estimate over are themselves functions of the summary statistic parameters, i.e. $t_i = f(x_i\;\varphi)$. The resulting density estimate $p(t|\varphi)$ will then be differentiable as long as the kernel $K$ is differentiable with respect to $t_i$, and by extension with respect to $\varphi$. To extend this differentiability in a binned fashion, we can accumulate the probability mass of the KDE within the bin edges of the original histogram -- equivalent to evaluations of the Gaussian cumulative density function --  to convert $p(t|\varphi)$ to a \textbf{binned KDE (bKDE)}, i.e. a set of discrete per-bin probabilities $p_i(\varphi)$.

\begin{figure}
    \centering
    \includegraphics[width=\textwidth]{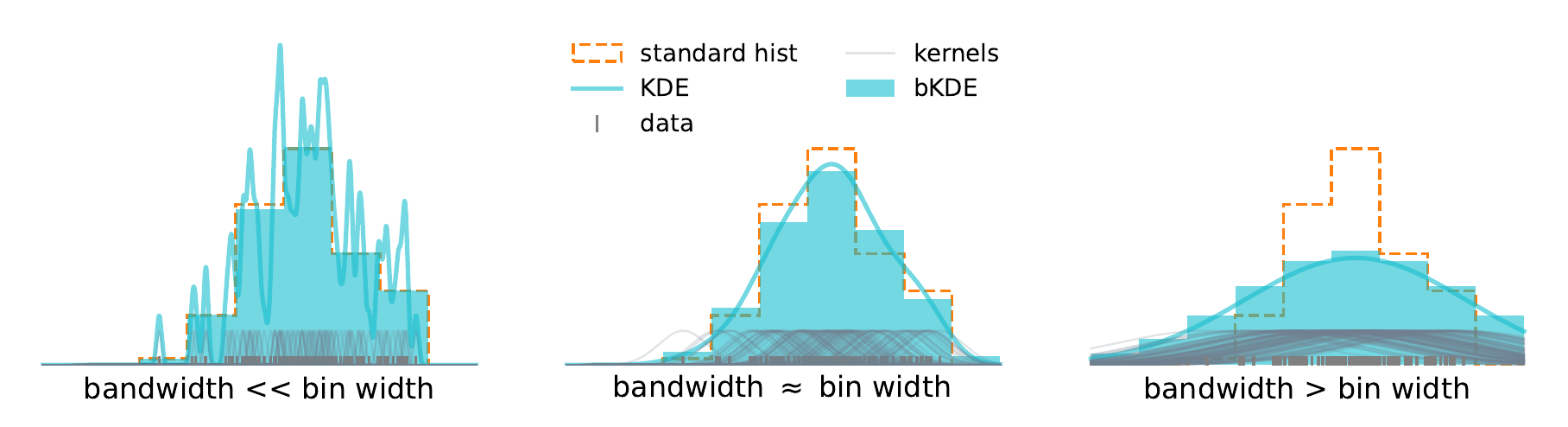}
    \vspace*{-8mm}
    \caption{Illustration of the bias/smoothness tradeoff when tuning the bandwidth of a bKDE, defined over 200 samples from a bi-modal Gaussian mixture. All distributions are normalised to unit area. The individual kernels that make up the KDE are scaled down for visibility.}
    \label{fig:hist}
\end{figure}

In the limit of vanishing bandwidth, the bKDE recovers the standard histogram, but gradients become susceptible to high variance. Increasing the bandwidth can alleviate this, but at the cost of introducing a bias. There is then a trade-off between decreasing the bandwidth enough to minimise this bias, and increasing it enough to guarantee gradient stability\footnote{How low is low enough when tuning? This relationship between bias and gradient stability is heavily impacted by the number of data points, and also by the width of the intervals, with more exploration of this planned.}. We can see a demonstration of this behaviour in \autoref{fig:hist}, where the bandwidth is tuned relative to the bin width.

\subsection{Likelihood modelling}

A popular framework to build statistical models for binned observations based on ``template" histogram data is HistFactory \cite{hifa}, used widely used across HEP. Thankfully, the resulting log-likelihood function $p_\mathbf{h}(x)$ is differentiable with respect to both the observed data $x$ and histogram data $\mathbf{h}$ from which the model is constructed. However, these gradients only recently became readily accessible in software, owing to the development of \texttt{pyhf} \cite{pyhf, pyhf2}: a Python package for building these likelihoods that leverages automatic differentiation.
\subsection{Test statistics (profile likelihood ratio)}

A number of test statistics are typically used in HEP\footnote{A full list of the different test statistics and their purposes can be found in \cite{asymptotics}.}, all of which build on the profile likelihood ratio. For likelihood function $p$, input data $x$, parameters of interest (POIs) $\mu$, and nuisance parameters $\psi$, we define the profile likelihood ratio as

\begin{align}
    \lambda_\mu(x) = \frac{p(x|\mu,\doublehat{\psi}(\mu))}{p(x|\hat{\mu},\hat{\psi})},
\end{align}
where $\hat{\mu},\hat{\psi}$ denote the maximum-likelihood parameters, whilst $\doublehat{\psi}(\mu)$ denotes the maximum-likelihood estimate of $\psi$ at fixed values of $\mu$. A test statistic $q_\mu$ is then built upon the profile likelihood, following prescriptions detailed in \cite{asymptotics}. Our physics objective, i.e. the expected $\mathrm{CL}_s$, is computed from distributions of $q_\mu$ for the null and alternative hypotheses, which correspond to choosing different values of $\mu$. In asymptotic theory, the null is distributed as a $\chi^2$-distribution, while the alternative follows a non-central $\chi^2_\mathrm{nc}(x,\Lambda^2)$ distribution, with $\Lambda^2$ given by 
\begin{equation}
\Lambda^2 = (\mu-\mu')^2/\sigma_{\hat{\mu}}^2,
\end{equation}
where $\mu$, $\mu'$ are the POI values corresponding to the null and alternative hypothesis respectively, and $\sigma_{\hat{\mu}}$ is the variance for the maximum likelihood estimate of the POIs. Our goal of minimising the expected $\mathrm{CL}_s$ is then equivalent to both maximising $\Lambda^2$ and minimising $\sigma_{\hat{\mu}}$. 

While the INFERNO approach uses the Fisher Information to estimate and minimise $\sigma_{\hat{\mu}}$, we take the route of computing $\Lambda^2$ as the test statistic value for the ``Asimov dataset'' of the alternative hypothesis. As described in~\cite{asymptotics}, the use of Asimov test statistics is known to yield better approximations of the true sampling distributions, and therefore of the final objective $\mathrm{CL_s}$. However, the drawback of this approach is the inclusion of optimisation routines such as the computation of $\doublehat{\psi}$, which are not a-priori differentiable in an efficient manner. We thus explore methods for derivatives of optimisation routines that only require gradient information when optimisation has converged.

\subsubsection{Implicit differentiation of optimisation routines:}

Optimisation algorithms find best-fit values $\hat{\theta}$ for a given objective function $f$. In the case of the profile likelihood, the objective $f$ is the likelihood function, which implicitly depends on the summary statistic parameters $\varphi$; we denote this by $f = f_\varphi$, and extend this dependence to the best-fit parameters $\hat{\theta}(\varphi)$. Iterative optimisation finds $\hat{\theta}$ by starting at a initial value $\theta_0$ and repeatedly updating it through a iteration step $\hat{\theta}_{i+1}=u(f_\varphi, \theta_i)$ until a terminating condition is satisfied. An explicit expression for $\hat{\theta}(\varphi)$ capturing the dependence on $\varphi$ is not readily available. Despite this, we can still compute gradients of $\hat{\theta}$ by noting that at the solution $\theta_i = \hat{\theta}$, the following relation holds for any $\varphi$:
\begin{equation}
    \hat{\theta} = u(f_\varphi,\hat{\theta}) ~\Rightarrow~ g(\varphi,\hat{\theta}) = u(f_\varphi,\hat{\theta}) - \hat{\theta} = 0,
\end{equation}
From here, we can leverage a powerful result from the \textit{implicit function theorem} that locally guarantees the existence of a continuously differentiable function $\hat{\theta}(\varphi)$, with gradients given by 
\begin{equation}\label{eq:fi}
    \frac{\partial\hat{\theta}(\varphi)}{\partial\varphi} = -\left[\frac{\partial g}{\partial \theta}\right]^{-1}  \frac{\partial g}{\partial \varphi} = \left[I - \frac{\partial u}{\partial \theta}\right]^{-1} \frac{\partial u}{\partial \varphi}
\end{equation}
This is our saving grace: by evaluating derivatives of\textit{ just the update step} with respect to the optimised parameters $\theta$ and the parameters implicitly defining the objective $\varphi$, we can derive gradients of the resulting optimisation solution without unrolling all iterations. Implicit differentiation through optimisation procedures like this has been implemented in software by the authors of the \texttt{jaxopt} package \cite{jaxopt}, which we use in our experiments. We refer the reader to Ref.~\cite{implicit} for more detail on the connection between \autoref{eq:fi} and automatic differentiation. 

This result lets us compute the non-centrality parameter $\Lambda^2$ of the hypothesis test (and as a corollary, the expected $\mathrm{CL_s}$) in a fully differentiable manner, providing the final building block of an end-to-end differentiable analysis pipeline as described in the beginning of this section.

\section{Demonstrating \texttt{neos}: End-to-end Analysis Optimisation}

We demonstrate the principle of end-to-end optimisation by training a neural network optimised directly on the physics sensitivity, expressed as the expected \CLs{} value of a \verb+pyhf+ model.

\subsection{Gaussian blob experiment}

\begin{figure}
    \centering
    \includegraphics[width=\textwidth]{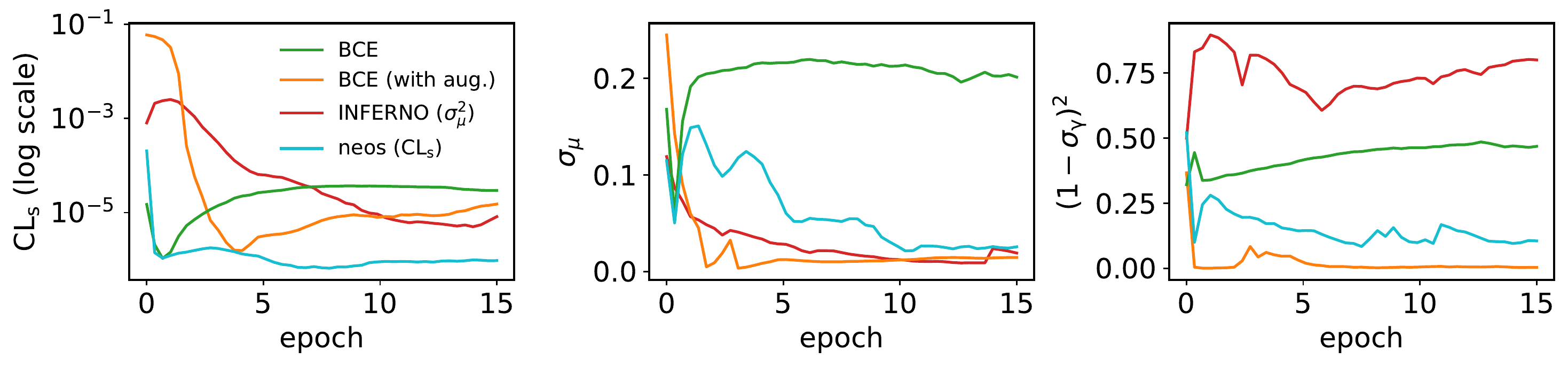}
    \vspace*{-8mm}
    \caption{Analysis metrics evaluated on a test set when running the \texttt{neos} pipeline with a variety of loss functions, averaged over 7 initialisations of the neural network weights. \textbf{Left}: the expected \CLs{} with a null hypothesis of no signal. \textbf{Middle}: the uncertainty on the signal strength $\mu$. \textbf{Right}: deviation of the nuisance parameter uncertainty $\gamma$ from its nominal value of 1.}
    \label{fig:metrics}
\end{figure}

We consider an example where we generate toy physics data $(x, y)$ from 2-D Gaussian blobs with covariances equal to the 2-D identity matrix $I_2$. 

The dataset $\mathcal{D}$ consists of \textcolor[rgb]{0.31,0.89,0.87}{signal} $\textcolor[rgb]{0.31,0.89,0.87}{s} \sim \mathcal{N}(m_s, I_2)$, \textcolor[rgb]{0.96,0.65,0.14}{background} $\textcolor[rgb]{0.96,0.65,0.14}{b} \sim \mathcal{N}(m_b, I_2)$, and two variations of the background data $\{\textcolor[rgb]{0.73,0,0.09}{b\mathrm{_{up}}}, \textcolor[rgb]{0.93,0.86,0}{b_{\mathrm{down}}}\}$ that mimic simulating the background process after varying an imagined physical parameter $\Psi$ \textcolor[rgb]{0.73,0,0.09}{up} and \textcolor[rgb]{0.93,0.86,0}{down} by one standard deviation, which is a common situation encountered in HEP. In practice, $\Psi$ is taken to influence the mean of the background distribution implicitly through generating blob data with different means, i.e. $\textcolor[rgb]{0.73,0,0.09}{b\mathrm{_{up}}} \sim \mathcal{N}(m_{b{\mathrm{up}}}, I_2)$, $\textcolor[rgb]{0.93,0.86,0}{b_{\mathrm{down}}} \sim \mathcal{N}(m_{b{\mathrm{down}}}, I_2)$.

After dataset generation, as well as splitting into train and test datasets, the workflow then follows the steps outlined in \autoref{fig:pipe}. Results from running this experiment can be found in \autoref{fig:metrics}, which compares \texttt{neos} to three baseline loss functions:
\begin{itemize}
    \item Binary cross-entropy (BCE): signal vs. nominal background
    \item BCE, but using up/down systematic variations as data augmentation for the background sample. This is a powerful baseline that maximally isolates the signal points from all background variations.
    \item INFERNO: using the signal strength uncertainty $\sigma_\mu^2$ as the objective, calculated from the Hessian of the likelihood via automatic differentiation.
\end{itemize}

Three metrics are shown: the expected \CLs{}, the parameter of interest ($\mu$) uncertainty, and the nuisance parameter ($\gamma$) uncertainty, all evaluated on the test set\footnote{During test evaluation and training with binary cross-entropy, no differentiable approximations are used.}. All metrics are averaged over seven different initialisations of the weights $\varphi$. Hyperparameters are given in \hyperref[app1]{Appendix A}.

We can see that \texttt{neos} is able to reach sensitivities lower than any other loss function (left plot). Moreover, it is able to gradually improve the uncertainty on the parameter of interest (middle plot) without any additional tuning or regularisation, far outperforming binary cross-entropy in this regard. When mixing in the up/down systematic variations as part of the background class when using BCE, we are able to get similar performance to \texttt{neos} with lower $\mu$ uncertainty around epoch 4, but \texttt{neos} is able to improve to match this with more training, all while retaining a lower \CLs{}. INFERNO is able to achieve the best uncertainty on $\mu$, but fails to reach the same expected sensitivity as other methods.

We additionally examine the uncertainty on $\gamma$ (right plot), also taken from the Fisher information matrix. It indicates whether the analysis is able to determine the systematic effect on the background shape in a way that differs from the constraint on $\gamma$ provided by the up/down variations. This is typically undesirable behaviour, as it indicates a strong dependence of the summary statistic on the nuisance parameters. Here, we see that \texttt{neos} and BCE (with aug.)  are able to keep this metric fairly low,  while BCE and INFERNO show signs of over- or under- constraining $\gamma$.

\section{Discussion and Future Directions}

One of the more immediate concerns for using \texttt{neos} in practice is the issue of scaling. Training times compared to BCE in the toy problem are a factor of 3 or so higher, since the processing of one batch corresponds to running the all the downstream parts of a physics analysis. In addition, the batch size needs to be large enough to faithfully represent the analysis, hopefully avoiding situations with empty bins. This is an issue that is currently being investigated; a medium/long term goal would be to scale \texttt{neos} in application to open data from the major LHC experiments, such as ATLAS and CMS, allowing the probing of how the performance benefits highlighted in section 3 scale to realistic problems with many sources of uncertainty.

Despite the focus on summary statistics, the work done in enabling the differentiability of HEP analysis opens up a variety of new approaches in optimising any free parameters end-to-end. For example, one could imagine optimising the pre-selection stage that occurs when filtering data prior to inference, since one can differentiably approximate a cut with a sigmoid function. Moreover, one could optimise the binning of an observable for the best sensitivity, given a fixed number of bins.

An additional, exploratory direction is considering the use of a multi-term objective function with weighted components. The software released alongside this work already enables objectives of the form  $a_0{\mathrm{CL_s}} + a_1{\sigma_\mu} + a_2{(1-\sigma_\gamma)^2} + \dots$, which could offer regularisation effects to steer away from pathologies that happen to satisfy one metric only. Work in this direction is also ongoing.

A future comparison to black box methods such as Bayesian optimisation would also be desirable, as they do not require the use of approximations to enable optimisation, but still can target the physics goals of choice.

\section{Conclusion}

The typical HEP analysis workflow has been made differentiable, and a use case for it in  the systematic-aware optimisation of a neural network has been demonstrated. Comparison was made to the more standard approach of using binary cross-entropy, as well as more competitive baselines like systematic data augmentation and INFERNO. Improvements were shown in both the expected analysis sensitivity and the properties of the likelihood function. Accompanying software to run the experiments detailed in Section 3 can be found at \cite{neos}, as well as a toolbox called \texttt{relaxed}~\cite{relaxed} that enables the isolated or combined use of the differentiable components used to assemble \texttt{neos}.

\ack
We would like to thank the authors of INFERNO (Pablo de Castro, Tommaso Dorigo) for initial productive discussions. We also extend thanks to Kyle Cranmer, Alexander Held, Giordon Stark, and Matthew Feickert for conceptual and technical support in the development of this project. Visualisations are produced using the \texttt{matplotlib} package \cite{mpl}. LH is supported by the Excellence Cluster ORIGINS, which is funded by the Deutsche Forschungsgemeinschaft (DFG, German Research Foundation) under Germany’s Excellence Strategy - EXC-2094-390783311. NS completed this work with financial support from Insights, which is funded by the European Union’s Horizon 2020 research and innovation programme, call H2020-MSCA-ITN-2017, under Grant Agreement n. 765710. NS extends gratitude to friends and colleagues that indirectly facilitated this work, particularly to those in the ``noonch'' lunch group at CERN.

\appendix
\section{}\label{app1}

Hyperparameters for the Gaussian blob study in Section 3:

\begin{itemize}
    \item 10000 data points, split evenly between all four blobs,
    \item 3-layer neural network of size (1024, 1024, 1),
    \item Training with Adam optimiser, learning rate 1e-3,
    \item Adam optimiser also used in maximum likelihood fits with learning rate 1e-2,
    \item $m_\mathrm{s}=(-1, 1)$, $m_\mathrm{b}=(2.5, 2)$, $m_\mathrm{bup}=(-2.5, -1.5)$, $m_\mathrm{bdown}=(1, -1)$,
    \item Multiplicative histogram scale factors: signal scale=2, background scale=10, global scale=10,
    \item ReLU activations, with sigmoid activation on the final layer,
    \item 15 epochs, with a batch size of 2000.
\end{itemize}

\section*{References}
\bibliographystyle{iopart-num}
\bibliography{bib}

\providecommand{\newblock}{}
\begin{thebibliography}{10}
\expandafter\ifx\csname url\endcsname\relax
  \def\url#1{{\tt #1}}\fi
\expandafter\ifx\csname urlprefix\endcsname\relax\def\urlprefix{URL }\fi
\providecommand{\eprint}[2][]{\url{#2}}

\bibitem{2012}
ATLAS-Collaboration 2012 {\em Physics Letters B\/} {\bf 716} 1–29 ISSN
  0370-2693 \urlprefix\url{http://dx.doi.org/10.1016/j.physletb.2012.08.020}

\bibitem{inferno}
de~Castro P and Dorigo T 2019 {\em Computer Physics Communications\/} {\bf 244}
  170–179 ISSN 0010-4655
  \urlprefix\url{http://dx.doi.org/10.1016/j.cpc.2019.06.007}

\bibitem{uncert}
Ghosh A, Nachman B and Whiteson D 2021 {\em Physical Review D\/} {\bf 104} ISSN
  2470-0029 \urlprefix\url{http://dx.doi.org/10.1103/PhysRevD.104.056026}

\bibitem{pivot}
Louppe G, Kagan M and Cranmer K 2017 Learning to pivot with adversarial
  networks (\textit{Preprint} \eprint{1611.01046})

\bibitem{hifa}
Cranmer K, Lewis G, Moneta L, Shibata A and Verkerke W (ROOT Collaboration)
  2012 {HistFactory: A tool for creating statistical models for use with RooFit
  and RooStats} Tech. rep. New York U. New York
  \urlprefix\url{https://cds.cern.ch/record/1456844}

\bibitem{pyhf}
Heinrich L, Feickert M and Stark G {pyhf: v0.6.3}
  https://github.com/scikit-hep/pyhf/releases/tag/v0.6.3
  \urlprefix\url{https://doi.org/10.5281/zenodo.1169739}

\bibitem{pyhf2}
Heinrich L, Feickert M, Stark G and Cranmer K 2021 {\em Journal of Open Source
  Software\/} {\bf 6} 2823 \urlprefix\url{https://doi.org/10.21105/joss.02823}

\bibitem{asymptotics}
Cowan G, Cranmer K, Gross E and Vitells O 2011 {\em The European Physical
  Journal C\/} {\bf 71} ISSN 1434-6052
  \urlprefix\url{http://dx.doi.org/10.1140/epjc/s10052-011-1554-0}

\bibitem{jaxopt}
Blondel M, Berthet Q, Cuturi M, Frostig R, Hoyer S, Llinares-L{\'o}pez F,
  Pedregosa F and Vert J~P 2021 {\em arXiv preprint arXiv:2105.15183\/}

\bibitem{implicit}
Duvenaud D, Johnson M and Kolter Z Deep implicit layers - neural odes, deep
  equilibirum models, and beyond
  \urlprefix\url{http://implicit-layers-tutorial.org/}

\bibitem{neos}
Simpson N and Heinrich L 2021 {neos: version 0.2.0}
  \urlprefix\url{https://github.com/gradhep/neos}

\bibitem{relaxed}
Simpson N 2022 {relaxed: version 0.1.3}
  \urlprefix\url{https://github.com/gradhep/relaxed}

\bibitem{mpl}
Hunter J~D 2007 {\em Computing in Science \& Engineering\/} {\bf 9} 90--95

\end{thebibliography}

\end{document}